# The role of unit systems in expressing and testing the laws of nature

Paul Quincey and Kathryn Burrows

National Physical Laboratory[1], Hampton Road, Teddington, TW11 0LW, United Kingdom

E-mail: paul.quincey@npl.co.uk

**Abstract**

The paper firstly argues from conservation principles that, when dealing with physics aside from elementary particle interactions, the number of naturally independent quantities, and hence the minimum number of base quantities within a unit system, is five. These can be, for example, mass, charge, length, time, and angle. It also highlights the benefits of expressing the laws of physics using equations that are invariant when the size of the chosen unit for any of these base quantities is changed. Following the pioneering work in this area by Buckingham, these are termed "complete" equations, in contrast with equations that require a specific unit to be used. Using complete equations is shown to remove much ambiguity and confusion, especially where angles are involved. As an example, some quantities relating to atomic frequencies are clarified. Also, the reduced Planck constant $\hbar$, as commonly used, is shown to represent two distinct quantities, one an action (energy x time), and the other an angular momentum (action / angle). There would be benefits in giving these two quantities different symbols. Lastly, the freedom to choose how base units are defined is shown to allow, in principle, measurements of changes over time to dimensional fundamental constants like $c$.

**Introduction**

The idea of a system of measurement units, whereby all measurement units are derived directly from a small number of base units, can be dated to Gauss in 1832, who used units for mass, length and time to measure magnetic fields. The formal debate about whether to add an electrical unit to the three mechanical units lasted some 47 years, from 1901 when proposed by Giorgi, to 1948 when the m-kg-s-A system was adopted. This, of course, formed the basis of the International System of Units (SI) in 1960. There has been an even longer debate about the analogous but less practically important status of angle within unit systems. Many papers over many decades have raised the issue; [1-3], for example. They have typically focussed on specific problems, and have proposed similarly specific solutions. We believe that by starting with the broadest possible context, which is the relationship between unit systems and the equations used to express the laws of nature, one of several benefits is that any problems relating to angle are easily resolved, at the expense only of inserting an unfamiliar constant into some equations in an entirely logical and consistent way.

**The case for five independent physical quantities, including angle**

An earlier paper [4] described how it is useful to consider unit systems such as SI, Gaussian and "natural unit" systems as being simplifications of a more general system with five independent, or base, units. It did not attempt to justify that number on physical grounds. Such a justification is presented here.

---

[1] The paper expresses the views of the authors and these do not necessarily reflect those of NPL Management Ltd.

The use of a system of units to ensure there is only one unit for each quantity – the energy example

As a preliminary point, it is helpful to consider the role of a unit system in ensuring that there is only one unit for each quantity, or in other words removing conversion factors from equations, numbers that simply convert from one unit to another for the same quantity. This is in addition to more pragmatic roles of providing a single system for global use, with simple power-of-ten multiples and sub-multiples of units to cover the required ranges, and of providing clear units for all relevant quantities.

The mechanism for eliminating conversion factors can best be seen by considering the various units that have been used for energy, such as the joule, erg, calorie, electron-volt, horsepower-hour and so on. The principle of conservation of energy, developed in the mid-19th century, included the realisation that many combinations of quantities, such as force x distance, mass x (velocity)$^2$, torque x angle, charge x potential difference, and heat capacity x temperature difference, produce the same quantity, namely energy. If the units for each of these quantities are chosen organically, it is inevitable that many different units for energy will arise, leading to conversion factors, such as 4.18 J/cal, appearing in equations.

A unit system can avoid conversion factors by following a simple strategy. If a single unit for energy is chosen, then for each pair of quantities producing energy, one is given an independently-defined unit; the other can then be derived, fixed in size by the combination of the energy unit and the defined unit within the pair. Using the examples given above, base units could be assigned for energy, length, mass, angle, charge and temperature, leading to equations free from conversion factors. There is a large choice of the 6 base quantities required by this example; apart from using the other quantity in each pair, the energy unit could be defined by creating base units for both quantities within one of the pairs, so that length, mass, angle, charge, temperature and voltage would be equally suitable. Where there are *n* distinct combinations of quantities producing the same quantity, *n+1* base quantities are required for a unit system that avoids having more than one unit for that quantity.

This line of reasoning could be used to define a minimum necessary number of base units only if it is clear how many distinct forms of energy there are. We think that the consideration of conservation principles provides a more objective approach.

Determining the minimum number of base units from conservation principles

A given conserved quantity must presumably be physically distinct from other conserved quantities, or the concept of conservation is meaningless. If the scope of the unit system excludes elementary particle interactions, where quantities such as colour charge and weak isospin are conserved, there are just four generally conserved quantities in physics: mass-energy, linear momentum, angular momentum, and electric charge. It should be emphasised that these conservation principles do not break down when elementary particle interactions are considered, they are instead supplemented by other ones, and also that they apply across a very broad range of physics, including relativistic and quantum physics[2].

In terms of a unit system, and its role of eliminating conversion factors, we must consider how these four distinct quantities form part of a common physical framework. The natural linkage is through their conjugate quantities and the quantity of action:

---

[2] It is tempting to think that because space and time form a single framework in relativistic physics, there is no relativistic distinction between length and time – a misconception that is reinforced by the habit of casually setting *c* = 1. However, space-time consists of 3 space axes and 1 time axis, not 1 space axis and 3 time axes, or 4 equivalent axes of "space-time", so the distinction between length and time remains. The Lorentz Transformation shows that lengths and time intervals, and, likewise, energies and momenta, are inextricably linked when changing from one inertial reference frame to another, but energy and momentum are always separately conserved within each reference frame. Similarly, quantum physics refines the conservation laws, such as by allowing the creation of pairs of virtual particles in line with the Uncertainty Principle, but the conservation laws remain. Energy, momentum, angular momentum and charge are separately conserved in beta decay, for example.

| | | | | |
|---|---|---|---|---|
| mass-energy | x | time | = | action |
| (linear) momentum | x | length | = | action |
| angular momentum | x | angle | = | action |
| electric charge | x | magnetic flux | = | action |

Following the same reasoning as in the previous section about energy, analogously ensuring that only one unit for action is created, we can see that the number of base quantities required for a unit system covering "practical" measurable quantities is five, for example energy, momentum, angular momentum, charge and action; or mass, length, time, angle and charge. We can understand this number as four distinctly conserved quantities plus one quantity that links them together, making five independent quantities in this sub-set of physics.

A superficially persuasive argument based on other simple physical principles can be made for four independent units: two to cover the distinct aspects of space-time, length and time, and two to cover the two non-nuclear forces, gravity and electromagnetism, for example mass and charge. However, this argument overlooks the distinction between the independently conserved quantities linear and angular momentum, and cannot account for the three independent quantities – mass, charge and angular momentum (spin) – possessed by elementary particles such as electrons.

From this point, we will use the term "coherent system of units" to mean one containing independently defined units corresponding to each of the five independent physical quantities[3]. Units for other quantities are derived from the base units in the usual way. The SI is an example of such a system, because it gives all five of these quantities independently defined units, even though angle is not treated as a base quantity within the system[4].

**The urge to reduce this number**

As described in more detail in [4], there is a natural tendency for theoreticians to simplify equations, removing dimensional constants from them by assigning new dimensions to some quantities and then setting the constants equal to 1.

This is most widespread in the treatment of angle, including within the SI. As described in more detail later, angle is generally treated as dimensionless, and the quantity of angle that we know as a radian is set equal to 1. This has the helpful consequence of simplifying equations that include angles, but the unhelpful consequence of excluding angles from dimensional analysis, creating confusion between frequency and angular velocity, and other issues that have always been "worked around".

Setting the velocity of light *c* equal to 1, as is done in most natural unit systems, is a more specialised example. There can be no issue with making the numerical value of this velocity equal to 1 by careful choice of the size of the units for length and time – for example using a light-year as the unit of length and the year as the unit of time. But to take the step from making *c* equal to 1 light-year per year, to making it equal to the number 1, and hence removing *c* from equations, can only be done at the expense of saying the dimensions of length are the same as those of time. According to Noether's theorem, this is equivalent to removing the distinction between energy and momentum, and hence removing one of the conservation laws. To give an example of why this is a problem, understanding elastic and inelastic collisions requires that energy is conserved and, separately, momentum is conserved. So while it is possible to simplify equations by setting dimensional quantities equal to one, effectively reducing the number of base quantities, in general this simplification comes at a high price. In this paper it is argued that the potential problems are avoided by understanding the difference between "complete" equations and "unit-specific" equations, as described below.

---

[3] This definition differs from that in JCGM 200:2012 only by requiring that the "given system of quantities" is the necessary set of five described above.

[4] The requirement for a set of "independently defined units" is that the units are defined in such a way that the size of each one of them can be changed without altering the size of the other units in the set. The SI base units are independent in this sense, even though changing the size of one unit may require numerical values within the definitions of other units to be changed, so that these other units remain the same size. Although it would not be sensible to redefine the size of a radian (the angle subtended at the centre of a circle by an arc that is equal in length to the radius), the radian is independent of the SI base unit definitions, and a different angle could be chosen as the SI unit, independently of the SI base units.

**The urge to increase this number**

The SI [5] of course contains seven base units (with the radian having unusual special status as a dimensionless derived unit). This can be seen as the recognition of the *usefulness* of labelling certain quantities as independent, in addition to the physical *necessity* of treating certain quantities as independent.

Temperature is the clearest case. The Boltzmann constant *k* can be set equal to 1 without oversimplifying the physics, because temperature can always be seen as a characteristic transferable energy, such as the average kinetic energy of gas molecules. In effect, temperature could be measured in joules without creating a problem with the physical dimensions. However, temperature is such a fundamental parameter of macroscopic systems that it is extremely useful to flag, with a dedicated unit, that a certain measure of energy is a temperature, and not some other quantity measured in units of energy.

**Different units for different quantities**

It has often been pointed out (for example in the SI brochure [5], p.140) that the SI usually ascribes different units to distinct quantities, but with notable exceptions, such as specific heat and entropy (both having the unit J/K); torque and energy (N m and J, which are the same thing); and action and angular momentum (J s and kg m$^2$ s$^{-1}$, which are the same thing).

However, if we consider that temperature has the underlying dimensions of energy, as above, and measure temperature in joules, heat capacity (the energy needing to be supplied to give a unit increase in characteristic energy), and entropy (disorder, e.g. log *W*) would be understandably dimensionless. These quantities are by nature a ratio and a logarithm of a count. There are many diverse quantities that are by nature ratios or counts (or logs of counts), such as strain and refractive index, and it is no surprise that they come without dimensional flags. These quantities need to be identified in other ways, such as by a clearer verbal description than is needed for dimensional quantities, where the unit provides a basic piece of information.

The shared units for the other pairs of quantities mentioned can be seen to be a consequence of removing angle from the set of dimensions [6]. If angle is treated as a base quantity, the unit for torque becomes, for example, J rad$^{-1}$, and that for angular momentum J s rad$^{-1}$, giving them their justified physical distinction from energy and action respectively. By analogy with the term "complete equation", described in the next section, these units could be called the "complete units" for torque and angular momentum.

These examples demonstrate that the system of five dimensions, each with an independent unit, brings tangible benefits in showing clear distinctions between different physical quantities, and hence in dimensional analysis. This is seen as an additional indication that such a system indeed reflects the underlying physics, rather than arbitrary human labelling.

**"Complete" equations; equations that are invariant in any coherent system of units**

It has rightly been stated, in various ways, that the sizes of measurement units are an arbitrary human choice, unlike the "laws of nature" which exist outside the human realm (e.g. [7]). It must be possible, therefore, to express these laws as equations that are independent of the choice of measurement units.

Nearly all familiar equations, such as $F = ma$, $F = kx$, $F = GMm/r^2$, $F = q_1q_2/4\pi\varepsilon_0 r^2$; $V = IR$, $R = \rho L/A$, $E = hf$, $E = mc^2$, $v = f\lambda$, $Q = mc\Delta T$, $PV = nRT$, $A = \pi r^2$, $N(t) = N_0 e^{-\lambda t}$ (where the symbols have their usual meanings), do not rely on SI units being used; indeed they were all discovered before the SI was adopted. This does not mean that any unit can be used for each quantity; but any coherent system of units (as defined above) can be used. Following the early work in this area by Buckingham [8] and Bridgman [9], we propose that these unit-invariant equations are termed "complete". They should not be called "SI" equations, because they are entirely general [10].

In contrast, consider this equation for the electrical force between two charges: $F = q_1q_2/r^2$, as used within the electrostatic and Gaussian versions of the CGS system. In this case, the unit for charge cannot be chosen independently of the units for mass, length and time, and it is therefore not a complete equation. Bridgman used the term "adequate" for such equations; we will here use the term "unit-specific".

It is important to note that unit-specific equations are not incorrect. Like equations that make use of "natural units", they are often convenient simplifications of complete equations, and their solutions are entirely valid. However, from a metrological point of view, the complete equations should always be taken as the starting point, for two reasons. Firstly, they provide absolute freedom to define and redefine the size of each base unit, rather than be forced to take any unit definition that is "built-in" to the equation. The ability to define and redefine the size of base units is at the heart of metrology, and can have profound consequences, as shown in the final sections of this paper. Secondly, they provide a transparent dimensional relationship between the quantities in the equation, which is missing from unit-specific equations. In the case of the Coulomb equation given above, you would deduce from the equation alone that charge has the dimensions of length multiplied by the square root of force, not a dimension independent of mass, length and time, as required by conservation laws. Taking unit-specific equations as the starting point for discussions about the dimensions and units for the quantities within them is guaranteed to lead to confusion or *ad hoc* reasoning.

Unit-specific equations using "natural units" will not generally include dimensional constants such as $h$ and $c$. Theoreticians sometimes imply that by avoiding such constants, and the arbitrary units needed to quantify them, such equations are closer to the fundamental physics. However, it would be more accurate to say that such equations *hide* the constants outside the equations rather than *avoid* the need for them, and that the complete equations show the physics in its most transparent form. Complete equations can be simplified to unit-specific equations by applying rules such as "set $c$ equal to 1", whereas the operation cannot be reversed simply by the rule "set 1 equal to $c$". In effect, unit-specific equations have lost some essential physical information compared with their complete versions.

**The equations containing angles that are in common use, and the need for the constant $\theta_N$**

Looking at familiar equations where angles are involved, such as $s = r\theta$ and $\omega = 2\pi f$, where again the symbols have their usual meanings, it is clear that they are only correct when the unit used for angle is the radian. If we wanted to measure angle in degrees, and angular velocity in degrees per second, for example, the equations are not correct. These are therefore unit-specific rather than complete equations, which should not be used to infer dimensional relationships.

It is tempting to suggest that angle is somehow fundamentally different from other measured quantities, and that the radian, as well as being the SI unit is somehow the only possible unit. However, we have seen earlier that angle can claim to have equal status with length and time on grounds of basic physics. Certainly the radian is the natural unit of angle, in the sense that its use simplifies equations, but that does not mean that the simplified equations are the complete, general versions, or that the radian is inherently dimensionless [11].

To be independent of the unit for angle, the complete equations for angular quantities need to include an unfamiliar constant, by analogy to the CGS-Gaussian equation for the force between two charges given above, which needs an extra constant (generally called $k_C$ or $1/4\pi\varepsilon_0$) to make it a complete equation. To our knowledge, this point was first made explicit by Brinsmade [1], who used "rad" for the constant we shall call $\theta_N$, and Torrens [2], who used $1/\eta$. $\theta_N$ is simply a specific quantity of angle[5], which could be defined as the angle at the centre of a circle for which the arc length equals the radius. It can be expressed in any angle units, and is, of course, equal to 1 radian or $180/\pi$ degrees, for example. As it is customary to name such constants after people, it would be appropriate to call this constant the Cotes angle, after Roger Cotes, who appreciated its significance in 1714. It is helpful to also give a symbol to the angle of one revolution, $\theta_{rev}$, which is $2\pi$ rad or 360 °, where $\theta_{rev} = 2\pi\theta_N$.

The complete equations for arc length are $s = r\,\theta/\theta_N = 2\pi r\,\theta/\theta_{rev}$; the second version emphasises that the angle $\theta$ simply fixes the arc length as a fraction of the circumference. The complete equations for angular frequency are $\omega = 2\pi\theta_N f = \theta_{rev} f$. The second version emphasises that the $2\pi$ in the first version is purely a numerical factor, which should not be interpreted as a number of radians. The frequency $f$ is expressed in reciprocal time units, while the angular frequency $\omega$ is expressed as angle units per time unit – clearly different quantities with different dimensions.

**The Radian Convention: setting $\theta_N$ equal to 1**

The current situation in the SI and elsewhere is simply that, by convention, $\theta_N$ is set equal to 1, in exactly the same way that $1/4\pi\varepsilon_0$ is set equal to 1 in some versions of CGS units, making the commonly-used equations involving angles unit-specific. This is discussed more fully in [4]. We call this the Radian Convention. It has become so widespread that most scientists consider that it is somehow part of the fabric of physics, rather than a convention that has been tacitly adopted and which can be simply reversed by replacing the "$\theta_N$"s that were, in effect, crossed out. The argument for five independent quantities, presented earlier, suggests that this is the only dimensional constant that has been tacitly set equal to 1 within the SI.

**Examples from wave theory and atomic physics**

The complete equations for frequencies and waves

The amplitude $\psi$ of a one-dimensional wave is typically described mathematically by:

$$\psi = \psi_0 e^{i(kx - \omega t)}$$

where $k = 2\pi/\lambda$ (unit e.g. m$^{-1}$) and $\omega = 2\pi f$ (unit e.g. s$^{-1}$)

All three of these equations implicitly adopt the Radian Convention. If it is removed, we have these complete equations, which are valid for any units for length, time and angle:

$$\psi = \psi_0 e^{i(kx - \omega t)/\theta_N}$$

where $k = \theta_{rev}/\lambda = 2\pi\theta_N/\lambda$ (unit e.g. rad/m) and $\omega = \theta_{rev} f = 2\pi\theta_N f$ (unit e.g. rad/s).

If metrological discussions involving waves and angular quantities start from the complete equations, all sources of potential confusion disappear. For example, including the Cotes angle in the equations allows angle to be included in dimensional analysis, a useful check on an equation's validity.

The two meanings of $\hbar$

We have seen that when the Radian Convention is removed, action and angular momentum become clearly distinct quantities, with units J s and J s rad$^{-1}$ respectively. The question then arises: which quantities do the symbols $h$ (the Planck constant) and $\hbar$ (the reduced Planck constant) represent?

Without labouring the point, the symbol $h$, in equations such as $p = h/\lambda$, is found to always represent an action. $\hbar$, on the other hand, represents an action equal to $h/2\pi$ in, for example, the Schrödinger equation (as in Table 2 below), but an angular momentum equal to $h/2\pi\theta_N$ in, for example, the equation $E = \hbar\omega$. Equations could be made complete by inserting $\theta_N$ into those where $\hbar$ represents an angular momentum. Instead, we propose making the minimal change to equations by using different symbols for the two meanings of $\hbar$.

---

[5] The subscript N is intended to suggest that the constant is the *natural* unit of angle.

There are then 3 main choices: 1) to keep $\hbar$ as the action $h/2\pi$, and have a new symbol for the angular momentum $h/2\pi\theta_N$; 2) to redefine $\hbar$ as the angular momentum $h/2\pi\theta_N$, and have a new symbol for the action $h/2\pi$; or 3) to redefine $\hbar$ as the angular momentum $h/2\pi\theta_N$, and have no symbol for $h/2\pi$, instead inserting $2\pi$ into equations as required. We propose option (2), with a new symbol $\check{h}$ for the action $h/2\pi$, as in Table 1, on the grounds that $\hbar$ is more often treated as an angular momentum than an action, especially in atomic physics, while the use of a special symbol for $h/2\pi$ is justified as it removes "$2\pi$"s from many equations.

Of course, when the Radian Convention is adopted, the units and numerical values for $\check{h}$ and $\hbar$ are indistinguishable, but we suggest that it would be very helpful to retain the different symbols even so.

| Symbol | Description | Pronounced in equations | Equal to | Value in selected complete units |
|---|---|---|---|---|
| | Angle | | | |
| $\theta_N$ | the Cotes angle | theta-N (or -nat) | | 1 rad, 57.3 ° |
| $\theta_{rev}$ | the angle of a revolution | theta-rev | $2\pi\theta_N$ | $2\pi$ rad, 360 ° |
| | Action | | | |
| $h$ | the Planck constant | h | (unchanged) | $6.63 \times 10^{-34}$ J s |
| $\check{h}$ | the reduced Planck constant | h-red | $h/2\pi$ (new symbol) | $1.05 \times 10^{-34}$ J s |
| | Angular momentum | | | |
| $\hbar$ | the Planck angular momentum | h-bar | $h/2\pi\theta_N = h/\theta_{rev} = \check{h}/\theta_N$ (new definition) | $1.05 \times 10^{-34}$ J s rad$^{-1}$ $1.84 \times 10^{-36}$ J s deg$^{-1}$ |

Table 1: Proposed symbols for specific quantities of angle, action and angular momentum

Acknowledging the difference between the two quantities $\check{h}$ (an action) and $\hbar$ (an angular momentum) would make the equations of quantum mechanics complete, and therefore dimensionally consistent, so that dimensional analysis could be applied with confidence. It would be good metrological practice, as discussed above, and would lead to reduced confusion, as we shall discuss in what follows later.

With $h$ and $\hbar$ defined as in Table 1, the equations $E = hf$, $p = h/\lambda$, $E = \hbar\omega$, $p = \hbar k$ and $F_z|\Psi\rangle = \hbar m_F|\Psi\rangle$, are complete, and do not change when the Radian Convention is adopted.

Other complete equations differ from their familiar Radian Convention versions. Some examples of such wave and quantum mechanical equations are shown in Table 2, again using the symbol definitions proposed in Table 1.

| Equation with the Radian Convention adopted (where $\theta_N = 1$ and $\check{h} = \hbar$) | Complete equation (using the symbols proposed in Table 1) |
|---|---|
| $\omega = 2\pi f$<br>$k = 2\pi/\lambda$<br><br>$\alpha = e^2/4\pi\varepsilon_0\hbar c \approx 1/137$<br><br>Bohr radius = $4\pi\varepsilon_0\hbar^2/m_e e^2$<br><br>$\psi = \psi_0 e^{i(px-Et)/\hbar}$<br><br>$-\dfrac{\hbar^2}{2m}\nabla^2\Psi + V\Psi = i\hbar\dfrac{d\Psi}{dt}$ | $\omega = \theta_{rev}f = 2\pi\theta_N f$<br>$k = \theta_{rev}/\lambda = 2\pi\theta_N/\lambda$<br><br>$\alpha = e^2/4\pi\varepsilon_0\check{h}c \approx 1/137$<br><br>Bohr radius = $4\pi\varepsilon_0\check{h}^2/m_e e^2$<br><br>$\psi = \psi_0 e^{i(px-Et)/\check{h}}$<br><br>$-\dfrac{\check{h}^2}{2m}\nabla^2\Psi + V\Psi = i\check{h}\dfrac{d\Psi}{dt}$ |

Table 2: Selected complete wave and quantum mechanical equations, together with their more familiar Radian Convention form, using the symbol definitions in Table 1.

The Rabi frequency

Quantum mechanics predicts that atom-light interactions with near-resonant coherent light cause the measurement probability to oscillate. The Rabi frequency is a measure of the periodicity of population occupancies within such atom-light interactions [12]. Students or those new to the field of atomic physics could be forgiven for thinking that the Rabi frequency is a frequency in the sense of $f = 1/T$, with the dimensions of reciprocal time. Its name, the fact that it is often expressed in Hz (it is not uncommon to see statements such as 'the Rabi frequency is 5 kHz'), as well as the fact that the Radian Convention does not differentiate dimensionally between $\omega$ and $f$ make it plausible to think that this is the case. This creates the practical problem that dimensional analysis cannot be used to assess whether or not a factor of $2\pi$ is required in equations.

The simplest case is that in which the atom is assumed to only have two possible states, usually labelled a ground state (g) and an excited state (e). The population probability ($|c(t)|^2$) that the atom is measured to be in the ground state is then given by:

$$|c_g(t)|^2 = \frac{1}{2}|c_g(0)|^2[1 + \cos(\Omega t)] + \frac{1}{2}|c_e(0)|^2[1 - \cos(\Omega t)]$$

It is clear from the above expression that the Rabi frequency, here denoted by $\Omega$, is an angular frequency, which should be expressed in the units of angle/time. This is known within the atomic physics community, such that when a value for the Rabi frequency of 5 kHz is quoted, it is expected that the numerical value used within equations is $2\pi \times 5000$. However, this system relies on expert knowledge and cannot be considered best practice.

The solution is, firstly, to use different symbols for the different quantities, let us say $\Omega_f$ and $\Omega_\omega$ for frequency (with unit e.g. s$^{-1}$) and angular frequency (e.g. rad/s) respectively, and, secondly, to start from complete equations. Here we will continue to use the symbol $\Omega$ for $\Omega_\omega$. The complete equations are:

$$\Omega = \Omega_\omega = \frac{g_F\mu_B B}{\hbar},$$

$$\Omega_f = \frac{\Omega}{\theta_{rev}} = \frac{\Omega}{2\pi\theta_N} = \frac{g_F\mu_B B}{2\pi\theta_N\hbar} = \frac{g_F\mu_B B}{h},$$

where $h$, $\hbar$, $\theta_N$ and $\theta_{rev}$ have the definitions given in Table 1, and the new symbols have their usual meanings.

The complete expression for population probability is[6]:

$$\left|c_g(t)\right|^2 = \frac{1}{2}\left|c_g(0)\right|^2 [1 + \cos(\Omega t/\theta_N)] + \frac{1}{2}|c_e(0)|^2[1 - \cos(\Omega t/\theta_N)]$$

Experimental values should be quoted either as, for example, $\Omega_f$ = 5 kHz or $\Omega$ = 2π x 5 krad/s. In this way, all confusion about equations and units is avoided, with the smallest possible deviation from current practices.

**The ability to redefine measurement units**

One of the powerful features of coherent systems of units, as we have defined them here, is the capacity to change the definitions of their base units while leaving the complete equations describing the world unchanged, and indeed leaving measurement results unchanged for practical purposes. This is usually done for reasons of increased precision, so that the accuracy of a measurement is not limited by the realisation of the unit. The redefinitions of the kg, A, K and mol in May 2019 are good examples of this [5].

At the heart of such changes is the idea that a unit represents a specific example of the relevant dimensional quantity, which exists independently of how it is defined. To take the redefinition of the metre in 1983 as an example, the move from a definition based on the wavelength of a Kr-86 emission, to one based on a frequency of a Cs-133 transition and a fixed speed of light, might seem to be a major change. But it was in reality a very pragmatic decision, enabling interferometers to use any laser wavelength as their basis without the accuracy being limited by an experimental value for the speed of light. The metre remained the same length for most practical purposes. The metre is also the same quantity, within experimental error, as the length of the metal metre bar in the Archives in Paris, and its original definition, a ten millionth of the distance between the North pole and the equator. The metre, like other specific dimensional quantities, has an objective size that can be quantified in many ways.

This capacity to redefine units contrasts with unit systems such as the Gaussian system, which sets $4π\varepsilon_0$ equal to 1, and systems of natural units, which additionally set other dimensional quantities equal to 1. For example, it is simply not possible for the Gaussian system to redefine its electrical unit by assigning a fixed value to $e$, as is now the case in the SI. Together with the fixed value for $h$, this brings consequent benefits for precision measurements of voltage and resistance arising from the quantities $2e/h$ and $h/e^2$ featuring in the Josephson and von Klitzing effects. Any fixed value for $e$ inevitably means that $4π\varepsilon_0$ can no longer be put exactly equal to 1.

It is the job of a good unit system to provide well-defined examples of dimensional quantities that are to be used as units, covering all necessarily (and usefully) independent physical quantities, which can be used as reference values for comparison with the quantities that are being measured. Of course, dimensional constants of nature such as $h$ and $e$, as used in the SI unit redefinitions of May 2019, provide a very natural and universal way of doing this. It should be mentioned that this use of natural dimensional constants to define the size of the units having the same dimensions is wholly different from treating them as "natural units" that are set equal to one, and hence considering them as dimensionless.

---

[6] Trigonometric functions such as cos ($x$) need extra care when angle units other than the radian are allowed. We suggest that, within equations in physics, the $x$ in cos ($x$) should always represent a dimensionless number, such that the cosine function can be expressed as the usual Taylor expansion in $x$, and the usual relations for differentiation and integration of trigonometric functions apply. Terms such as cos ($\theta$) and cos ($\omega t$), where $\theta$ and $\omega t$ represent angles, agree with this definition when the Radian Convention is adopted. Within complete equations, these terms must be replaced with cos ($\theta/\theta_N$) and cos ($\omega t/\theta_N$) respectively.

The use of fundamental constants to define the size of the SI base units is not expected to change in the future. However, this does not mean that we cannot use non-SI unit definitions for special purposes, as described in the next section.

**Dimensional constants and their change over time**

Many people have stated that it is not meaningful to discuss whether dimensional constants are varying over time, because, unlike dimensionless constants such as the fine structure constant $α$, they cannot be deconvolved from a man-made unit system, so that it would never be clear if it was the constant or the units that were changing in size (e.g. [7], [13-16]). Certainly it is possible to determine a change in $α$ without any direct consideration of measurement units, via the electron g-factor, for example, making the question considerably simpler. However, ruling out any meaningful change in dimensional constants goes against the idea that dimensional quantities exist independently of how we quantify them.

The situation is well illustrated by the question of whether it is possible to determine whether $c$ is changing over time. The key question is: compared to what? Moreover, a measurement of $c$ in SI units cannot possibly show a change – $c$ is 299 792 458 m/s by definition – which does indeed appear to indicate that any attempt to measure a change in $c$ is pointless.

However, we could choose, without changing the size of the length and time units in a unit system, just making them less precise, to define them using reference quantities that are, in principle, independent of $c$. These "Newtonian" references would need to be unaffected by changes in $c$, or, for our later convenience, in $h$, $e$ or $ε_0$. The size of an atom, and hence of a metal artefact or the Earth, would of course be strongly dependent on $h$, $e$ and $ε_0$, from simple consideration of the Bohr radius (shown in Table 2).

The motions of the planets in the solar system, on the other hand, are very well described by Newtonian mechanics and gravity, with negligible dependence on $h$, $e$, $ε_0$ or $c$ (providing $c$ stays much larger than the speeds of the planets[7]). If we define the time unit in terms of an astronomical year, as was the case in the SI between 1956 and 1967, and the length unit in terms of the distance between the Earth and the Sun (the Astronomical Unit), and measure $c$ after calibrating our instruments according to these definitions[8], a change in $c$ as measured in these "Newtonian" units would be apparent, irrespective of changes in $h$, $e$ or $ε_0$.

It is interesting to note that the first accurate measurement of $c$, by Bradley, was, in effect, made in these units. His method was to determine the ratio of $c$ to the velocity of the Earth in its orbit around the Sun, via the phenomenon of stellar aberration [17]. The constant of aberration is an experimentally determined angle of about 99 μrad (20.5 arcseconds). 2π divided by the sine of this angle provides a direct measure of the speed of light in Astronomical Units per year. Experimental determinations over time would therefore, in principle, show a change in $c$ irrespective of any change in $h$, $e$ or $ε_0$, and hence $α$. If $c$ and $h$ both changed in tandem, leaving their product the same, there would be observable consequences even though $α$ remained unchanged[9]. There is no point trying to

---

[7] Of course our best theory of gravity, general relativity (GR), is intimately linked to space-time, with space and time measures being related by the factor $c$. This suggests that the orbits of the planets would change if $c$ changed, when GR is considered. However, GR simplifies to Newtonian gravity when gravity is weak, and the speeds of objects are much less than $c$. The equations of Newtonian gravity hold when these conditions are met, whatever the value of $c$.

[8] The length unit could be made traceable to the Astronomical Unit using Cassini's 1672 method of measuring the size of the solar system using parallax, for example.

[9] Curiously, although these "Newtonian" units for length and time would vary with changes in $G$, simple physics shows that they would both vary in direct proportion to $G$, leaving the velocity of the Earth, and hence the constant of aberration, also unchanged, to first order, by changes in $G$. The constant of aberration would not be independent of all "fundamental constants" except $c$, however. If we assume that the masses of the Sun and

measure this change in *c* if your unit system states that *c* = 299 792 458 m/s, or *c* = 1, by definition. It is the ability of a coherent unit system to choose the definitions of its base units for specific purposes that makes it possible.

**Conclusions**

A great deal of time and energy spent discussing unit systems could be saved by, firstly, appreciating that they should be built on foundations of basic physics, not mathematics or arbitrary choices, and secondly, recognising the difference between complete equations and unit-specific equations. Basic physics tells us that, outside the realms of nuclei and elementary particles, five independently-defined units are necessary, and complete equations must therefore be invariant with changes to the five respective arbitrarily-sized units.

There would also be great educational value in always introducing the complete equation for any relationship, so that all dimensional constants are explicit, and dimensional analysis can be applied with confidence. Unit-specific equations can still be widely used, but with the understanding that convenient simplifications have been made to the complete equation, and dimensional analysis can then no longer be applied.

Complete equations are particularly beneficial within software that handles physical quantities and the relationships between them. Angular quantities need special treatment when the Radian Convention is adopted, but with complete equations they can be included in exactly the same way as any other dimensional quantity.

It is not the role of SI committees to tell scientists which equations to use in their work. However, it should be their role to explain that not all commonly-used equations are complete, such as those where the Radian Convention, CGS conventions, or natural units have been adopted, and that dimensional analysis of these unit-specific (i.e. non-complete) equations produces nonsense. It would also be helpful to point out that the "SI" equations for electromagnetism, for example, are the general, complete equations, and they do not rely on SI units being used.

It is similarly not the role of SI committees to tell scientists which symbols to use. However, it would also be helpful to point out that some quantities that are physically distinct, a status confirmed when complete equations are used, can appear to be the same thing when unit-specific equations are used. For example, the reduced Planck constant $\hbar$ and the Rabi frequency $\Omega$ are currently used in physically ambiguous ways, and this ambiguity can be removed by using different symbols for the distinct quantities.

Arguably it *is* the role of SI committees to tell scientists which units to use. It should at least be their role to point out that the complete SI units for angular momentum and torque are J s rad$^{-1}$ and J rad$^{-1}$ respectively, even though there is a long tradition of following the Radian Convention and using the units J s and N m instead.

Everyone can agree that the sizes of the units that we choose to use for practical measurements are arbitrary in physical terms. However, this should not lead us to underestimate the role of unit systems both in expressing the laws of nature in the most transparent way, as complete equations, and in finding ways to test them. In particular, it is clearly convenient to use dimensional constants of nature to define our measurement units, so that the constants have fixed values when these unit definitions are used. However, we should be aware that other unit definitions can be chosen, even within the same unit system, and this means that we have not lost the ability to test how constant the constants really are.

---

Earth scale in proportion to the mass of a proton, $m_p$, we would expect v/c in these units to scale with changes in $m_p$ as $m_p^{-1/2}$. It has also been assumed that the size of the angle measurement unit does not change over time, for example through a change in $\theta_N$.

**Acknowledgements**

We would like to thank Andrew Lewis for very helpful discussions.